\documentclass[conference]{IEEEtran}
\IEEEoverridecommandlockouts
\usepackage{cite}
\usepackage{amsmath,amssymb,amsfonts}
\usepackage{graphicx}
\usepackage{textcomp}
\usepackage{graphicx}
\usepackage[affil-it]{authblk} 
\usepackage{xcolor}
\usepackage{booktabs}
\def\BibTeX{{\rm B\kern-.05em{\sc i\kern-.025em b}\kern-.08em
    T\kern-.1667em\lower.7ex\hbox{E}\kern-.125emX}}

\usepackage{cite}
\usepackage{subcaption} % Assurez-vous d'inclure ce package dans le préambule
\usepackage{amsmath,amssymb,amsfonts}
\usepackage{graphicx}
\usepackage{booktabs} % Assurez-vous d'inclure ce 

\usepackage{tabularray}
\usepackage{textcomp}
\usepackage{tikz}
\usepackage{amsmath}
\usepackage{amsfonts}
\usepackage{adjustbox}
\usepackage{multirow}
\usepackage{xcolor}
\usepackage{pgfplots}
\usepackage{subcaption}
\usepackage{comment}
\usepackage[affil-it]{authblk} % Pour les affiliations
\usepackage{booktabs}
\usepackage{algpseudocode}
\usepackage[ruled, lined, longend, linesnumbered]{algorithm2e}
\usepackage[utf8]{inputenc}
\usepackage{tikz}

\usepackage{amssymb}% http://ctan.org/pkg/amssymb
\usepackage{pifont}% http://ctan.org/pkg/pifont

\usepackage{pgfplots}
\pgfplotsset{compat=1.18}
\usepgfplotslibrary{groupplots}
\usepackage{float}
\usepackage{comment} % Add this in the preamble

\begin{document}

\title{Fuse and Federate: Enhancing EV Charging Station Security with Multimodal Fusion and Federated Learning}

\author[1]{Rabah Rahal}
\author[1,2]{Abdelaziz Amara korba}
\author[2]{Yacine Ghamri-Doudane}

\affil[1]{Networks and Systems Laboratory (LRS), Badji Mokhtar Annaba University, Algeria}
\affil[2]{L3I, University of La Rochelle, France}

\maketitle

\begin{abstract}
The rapid global adoption of electric vehicles (EVs) has established electric vehicle supply equipment (EVSE) as a critical component of smart grid infrastructure. While essential for ensuring reliable energy delivery and accessibility, EVSE systems face significant cybersecurity challenges, including network reconnaissance, backdoor intrusions, and distributed denial-of-service (DDoS) attacks. These emerging threats, driven by the interconnected and autonomous nature of EVSE, require innovative and adaptive security mechanisms that go beyond traditional intrusion detection systems (IDS). Existing approaches, whether network-based or host-based, often fail to detect sophisticated and targeted attacks specifically crafted to exploit new vulnerabilities in EVSE infrastructure. This paper proposes a novel intrusion detection framework that leverages multimodal data sources, including network traffic and kernel events, to identify complex attack patterns. The framework employs a distributed learning approach, enabling collaborative intelligence across EVSE stations while preserving data privacy through federated learning. Experimental results demonstrate that the proposed framework outperforms existing solutions, achieving a detection rate above 98\% and a precision rate exceeding 97\% in decentralized environments. This solution addresses the evolving challenges of EVSE security, offering a scalable and privacy-preserving response to advanced cyber threats.

\end{abstract}

\begin{IEEEkeywords}
Electric Vehicle Supply Equipment (EVSE), Intrusion Detection System (IDS), Federated Learning, Cybersecurity, Multimodal Data Fusion
\end{IEEEkeywords}

\bibliographystyle{IEEEtran}

\section{Introduction}

% Context 
The global electric vehicle (EV) market has surged from 26 million EVs on the road in 2022 to over 40 million in 2024, reflecting a 53.85\% growth rate in sales \cite{1}. This rapid expansion makes it imperative to adapt current smart grid infrastructure, including Electric Vehicle Supply Equipment (EVSE), to meet the demands of this fast-growing sector. Each EVSE deployed must provide flexible payment solutions for access to services, such as credit card payments, mobile applications, or prepaid monthly cards. However, this accessibility to sensitive payment and user information has made EVSEs an attractive target for cyber threats, emphasizing the need for robust cybersecurity measures to protect user confidentiality and privacy.
 
% Sscurity issue and attaack againsta charing station 
Despite advances in securing traditional infrastructure, EVSEs continue to exhibit significant vulnerabilities. Recent incidents underscore the associated risks: in 2024, ransomware attacks on charging stations surged by 90\%, with cybercriminals encrypting station systems and demanding ransom payments for restoration \cite{cnews2024}. Furthermore, research has identified critical flaws in the widely-used Open Charge Point Protocol (OCPP), making charging sessions vulnerable to disruptions and unauthorized data access \cite{numerama2023, diaf2024beyond}. These incidents emphasize the urgent need for advanced cybersecurity solutions to protect EVSEs from a wide array of threats. Current security mechanisms, such as network-based \cite{diaf2024beyond, 10592500} and host-based IDS, often rely on modeling the behavior of EV charging stations using a single type of log within their local scope. This approach may lack the depth and adaptability required to detect sophisticated attack patterns and multi-layered cyber threats targeting EVSEs.

Current intrusion detection systems (IDS) for EV charging infrastructure have made progress in addressing cyber threats but face significant limitations. Many rely on complex architectures or multiple sub-models, which can be resource-intensive and unsuitable for deployment on constrained EVSEs \cite{3,4}. Others focus on specific attack scenarios like injection attacks in vehicle-to-grid communication but lack flexibility for diverse threats \cite{5,6}. Emerging AI-driven solutions incorporating blockchain and reinforcement learning (RL) show promise - blockchain ensures transaction security \cite{b1, b2} despite interoperability and storage challenges, while RL enables dynamic adaptation \cite{rl1, rl2} though requiring extensive data and raising ethical concerns. However, these advanced approaches still struggle with high computational costs, poor cross-network generalization, and the fundamental trade-off between scalability, adaptability, and resource efficiency. This highlights the critical need for more robust IDS frameworks capable of leveraging diverse data sources while overcoming these persistent limitations.

Developing a robust IDS for EVSEs is crucial to address cyberattacks like vulnerability scanning, network intrusions, and host-targeted attacks. Existing IDS solutions often focus on a single data source, such as network traffic \cite{korba2024ai,korba2023federated,tellache2024multi,10496859} or kernel logs \cite{10439152}, offering limited threat insight. Network traffic lacks host-level visibility, while kernel logs miss external attack patterns. Centralizing EVSE logs for global IDS training raises significant privacy concerns.

To address these challenges, our framework employs a comprehensive approach by intelligently processing and fusing diverse log types generated by EVSEs during operation. This enables the IDS to correlate information across multiple data sources, uncovering complex attack patterns that cannot be detected by analyzing a single log type alone. 

\begin{itemize}
    \item \textbf{Multimodal Data Fusion}: Our approach utilizes diverse data types generated by EVSEs, including network traffic and kernel events, to construct a comprehensive view of station activity. While this paper focuses on these two data sources, the framework's design, leveraging latent representation extraction and compression, is inherently extensible to incorporate other types of data logs, such as power consumption and voltage metrics.
    
    \item \textbf{Privacy-Preserving Collaboration}: Through Federated Learning (FL), the framework capitalizes on the diversity of logs collected by a large number of geographically distributed EVSEs, enabling collaborative model training without compromising data privacy.
    
    \item \textbf{Decentralized Security}: This decentralized approach ensures high detection accuracy while preserving the confidentiality of user and operational data.
\end{itemize}

By combining multimodal data fusion and federated learning, the framework significantly improves the detection of sophisticated attack patterns, providing a robust, scalable, and privacy-preserving cybersecurity solution for EVSE infrastructure.

The remainder of this paper is organized as follows: Section 2 reviews background and literature review; Section 3 defines the problem and model framework; Section 4 introduces the proposed methodology; Section 5 presents experiments and results; and Section 6 concludes with future research directions.

\section{Background and Related Work}

This section reviews EVSE foundations, focusing on key features, protocols, attack scenarios, and recent intrusion detection research.

\subsection{EVSE Characteristics and Communication Protocols}

The Electric Vehicle Supply Equipment (EVSE) ecosystem depends on key communication protocols for interoperability, security, and efficiency. Two main protocols are central to this: the Open Charge Point Protocol (OCPP) and ISO 15118.

OCPP, developed by the Open Charge Alliance, is a widely adopted protocol for managing communication between EV chargers and central systems. It facilitates features such as smart charging, remote diagnostics, and enhanced security, ensuring compatibility and centralized control across diverse EVSE networks \cite{7}. On the other hand, ISO 15118, an international standard for EVs, governs communication between vehicles and charging stations. This protocol introduces Plug \& Charge functionality, allowing automatic vehicle authentication, which improves user convenience and security. Additionally, ISO 15118 supports bidirectional energy transfer, enabling advanced vehicle-to-grid (V2G) services that enhance grid stability and energy efficiency \cite{8}.

\begin{figure*}[!t]
    \centering
    \vspace{0.2cm}  % Top padding
    \includegraphics[width=0.72\linewidth]{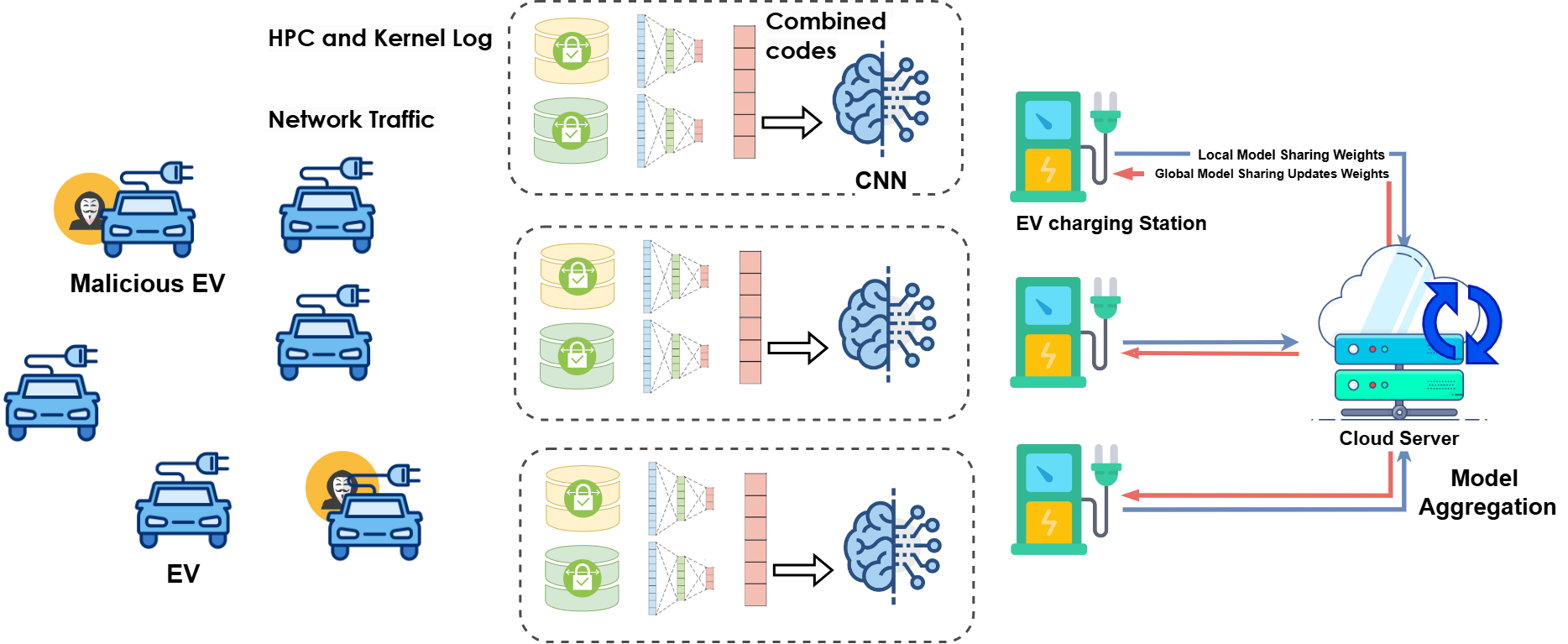}
    \caption{Federated Training of the Intrusion Detection Model}
    \label{fig:workflow}
\end{figure*}

\subsection{Related Work}\label{AA}

In \cite{3}, the authors proposed an IDS using ensemble learning for EVSE networks, targeting attack detection in both centralized and decentralized infrastructures. Their framework combines multiple models trained on diverse EVSE data to improve detection accuracy. However, relying on three classification sub-models may not suit all EVSEs, particularly those with limited computational resources.

Researchers in \cite{4} proposed an IDS for IoT-based EV charging stations to detect cyber threats, using a real IoT dataset with both binary and multiclass classification. They apply PCA for feature selection and evaluate performance with a CNN-A-LSTM model. However, the complex CNN-A-LSTM architecture poses risks of overfitting, highlighting potential limitations in generalization.

In \cite{5}, the authors present an injection attack on vehicle-to-grid (V2G) communication at public EV charging stations, exploiting ISO 15118 protocol flaws. Their custom testbed simulates malicious packet injections between the EV communication controller (EVCC) and the station’s Supply Equipment Communication Controller (SECC), leading to denial-of-service, remote code execution, and malware spread. They propose a machine learning-based IDS trained on benign and malicious traffic. However, the 64KB message size limit may reduce realism and hinder detection of complex threats, affecting IDS effectiveness.

Poudel et al. \cite{6} analyzed malware propagation in V2G communications at EVSEs, demonstrating malicious traffic injection risks. Their model maps EVSE connectivity to calculate malware spread probabilities, optimizing attacks for high-risk locations. Tests in urban/rural areas show optimal strategies increase spread by 10–33\%, revealing critical power-transportation vulnerabilities.

Several studies \cite{bh1, bh2, bh3} have developed anomaly detection systems to monitor EV charging behavior. While innovative, these approaches have some key blind spots. For example, \cite{bh1}'s clustering method identifies typical charging profiles well, but because it relies solely on historical data, it fails to detect sophisticated attacks where hackers mimic normal charging patterns. Behavior-based detection also faces challenges – without massive, diverse datasets, these systems often struggle to distinguish true threats from natural variations in EVSE usage.

Even though EVSE security has been enhanced by contemporary IDS solutions, edge deployment may not be feasible due to their weight. Our approach addresses this by combining host and network data in a smart and efficient way. Federated learning allows chargers to collaborate to improve detection while ensuring that training remains local (no data sharing is required), which is perfect for real-world EVSEs with constrained resources and ever-evolving threats.

\section{Proposed Solution}

\subsection{Overview}
We propose an intrusion detection framework for EVSE stations that leverages multiple log types, including network traffic and kernel events, to detect potential attacks. Operating locally on each EVSE as a host-based, multimodal system, the IDS uses feature extraction and an autoencoder to create compressed embeddings, which are fused to manage data dimensionality. This multimodal data fusion is integrated into federated training, where a cloud server acts as a parameter server and edge servers at charging stations function as clients, collaboratively training the detection model. By analyzing threats from multiple perspectives. 
Importantly, the overall framework is designed to be lightweight. It relies on a single CNN model and compact latent representations, which significantly reduces inference-time computation. In contrast to late fusion multimodal approaches that require multiple parallel models, our method is more energy-efficient and better suited for resource-constrained edge environments. Figure \ref{fig:workflow} illustrates the federated learning framework, with subsequent sections detailing the system architecture and training process.

\subsection{Feature Encoding and Fusion with Autoencoders}

Our system relies on two key types of data logs, each offering insights from a different perspective. The first, Network Traffic Logs ($D_1$), captures real-time communications between the charging station and external entities, such as payment gateways and central servers. Anomalies in network traffic, such as unusual spikes or abnormal request patterns, can indicate potential attacks. The second, Kernel Event Logs and High Performance Counter (HPC) Metrics ($D_2$), includes low-level kernel event logs and hardware performance metrics, providing a system-level view of charging station operations. These records are essential for detecting covert activities, such as privilege escalations or backdoor installations. For each dataset, relevant features are extracted based on statistical and domain-specific criteria, resulting in a feature vector $x_i^{(j)}$ for each dataset $D_j$, where $i$ denotes the station. Formally, the transformation can be represented as:

\begin{equation}
    x_i^{(j)} = f_j(\text{preprocess}(D_j)),
\end{equation}

where $f_j$ is the specific feature extraction function applied to each dataset.

%Given the unique nature of each data source, we apply a dedicated autoencoder to each dataset, allowing it to capture meaningful representations within a reduced latent space. Autoencoders, consisting of an encoder-decoder structure, compress the input into a lower-dimensional representation that retains significant information. 
We apply a dedicated autoencoder to each dataset to learn meaningful patterns in a compact latent space. The encoder-decoder structure compresses the input while preserving essential features. For each feature vector $x_i^{(j)}$ extracted from dataset $D_j$, we apply an autoencoder $g_j$ to obtain a latent representation $z_i^{(j)}$:

\begin{equation}
    z_i^{(j)} = g_j(x_i^{(j)}), \quad j \in \{1, 2\},
\end{equation}

where $g_j$ denotes the encoding function for each dataset $D_j$, preserving each modality's unique features in a compressed form for efficient storage and processing.

We adopt intermediate fusion by combining latent representations $z_i^{(1)}$ and $z_i^{(2)}$ from different data sources into a unified vector $z_i$, capturing the station’s overall state. Unlike early fusion, which merges raw inputs and may lack semantic depth, this approach integrates multimodal insights to enhance prediction quality. The combined vector $z_i$ is defined as:

\begin{equation} z_i = f_{\text{concat}}(z_i^{(1)}, z_i^{(2)}), \end{equation}

where $f_{\text{concat}}$ is the concatenation function, merging network and system features into a shared latent space.

\subsection{Attacks Detection}

For intrusion detection on fused vector $z_i$, we employ a 1D CNN, effective at capturing sequential/spatial patterns in latent features to identify malicious anomalies.

The CNN processes $z_i$ through multiple layers, each with distinct functions. The initial layers of the CNN apply convolutions to $z_i$, extracting critical feature patterns. For a convolutional filter $k$ of size $M$, the convolution operation on $z_i$ produces an output $c_i^{(k)}$ as follows:

\begin{equation}
    c_i^{(k)} = \sigma\left( \sum_{m=1}^{M} w_m^{(k)} z_{i,m} + b^{(k)} \right),
\end{equation}

where $w^{(k)}$ and $b^{(k)}$ are the filter's weights and biases, and $\sigma$ is a non-linear activation function, such as ReLU. Convolutional outputs pass through pooling layers to reduce dimensionality and capture key features:

\begin{equation}
    p_i^{(k)} = \max(c_i^{(k)}),
\end{equation}

where $p_i{(k)}$ represents the max-pooled value for filter $k$, Emphasizing prominent features while preserving spatial information, a softmax function outputs the intrusion probability:

\begin{equation}
    \hat{y}_i = \text{softmax}(W \cdot p_i + b),
\end{equation}

where $W$ and $b$ are the weights and bias of the final classification layer.

\begin{algorithm}
\caption{FED MDF-Based Attack Detection}
\label{algo1}

\KwIn{Multimodal datasets $\{D_1, D_2, \dots, D_M\}$ for each station $i$, with $N$ stations}
\KwOut{Intrusion detection probabilities $\{\hat{y}_i\}$ for each station $i$}

\For{each station $i = 1$ to $N$}{
    \For{$j = 1$ to $M$}{
        $x_i^{(j)} \gets f_j(\text{preprocess}(D_j))$ \tcp*[l]{Extract features from $D_j$}
    }
    
    \For{$j = 1$ to $M$}{
        $z_i^{(j)} \gets g_j(x_i^{(j)})$ \tcp*[l]{Encode features into latent space}
    }
    
    $z_i \gets f_{\text{concat}}(z_i^{(1)}, z_i^{(2)}, \dots, z_i^{(M)})$ \tcp*[l]{Combine latent vectors}
    
    \textbf{Initialize:} CNN parameters $\theta_i$ for station $i$\;
    $p_i \gets \text{CNN}_\text{1D}(z_i; \theta_i)$ \tcp*[l]{Process with 1-D CNN to extract intrusion features}
    $\hat{y}_i \gets \text{softmax}(W \cdot p_i + b)$ \tcp*[l]{Calculate intrusion probability}

    Compute gradients: $\nabla_{\theta_i} \mathcal{L}(h(z_i; \theta_i), y_i)$\;
}

\textbf{Federated Aggregation:} Update global CNN parameters\;
$\theta_{t+1} \gets \theta_t - \eta \sum_{i=1}^{N} \nabla_{\theta_i} \mathcal{L}(h(z_i; \theta_t), y_i)$\;

\Return Intrusion probabilities $\{\hat{y}_i\}$ for each station $i$\;
\end{algorithm}

To maintain data privacy, the CNN is trained using federated learning. Each charging station trains its local model on the fused vector $z_i$, updating the model parameters locally. Only the gradients or parameter updates are shared with a central server, which aggregates these updates to improve the global model without accessing raw data. The global parameter update is formulated as:

\begin{equation}
    \theta_{t+1} = \theta_t - \eta \sum_{i=1}^{N} \nabla_{\theta} \mathcal{L}(h(z_i; \theta_t), y_i),
\end{equation}

where:
$\theta_t$ represents the CNN parameters at iteration $t$,
$\eta$ is the learning rate, and
$\mathcal{L}$ is the local loss function, such as cross-entropy, computed at each station $i$.

The federated approach enables distributed training while preserving privacy and maintaining CNN detection effectiveness across stations. Algorithm \ref{algo1} outlines the smart contract's on-chain aggregation.

\section{Experiments and Results}

This section outlines the experimental setup and testing scenarios for evaluating separate/fused models and centralized/federated approaches in our framework.

%****************************************************

\subsection{Datasets}

In this research, we used CICEVSE2024 \cite{EVSE}, which is one of the newest publicly available datasets containing both benign and malicious traffic generated from a realistic testbed for Electric Vehicle Supply Equipment (EVSE). It includes data from three sources: Power consumption, Network Traffic, and HPC/Kernel Events (Table \ref{tab:distSamp}). The testbed involved an operational Level 2 charging station (EVSE-A), Raspberry Pi devices for various system components (EVCC, EVSE-B, Power Monitor, Charging Station Management System (CSMS)), and communication via OCPP and ISO15118 protocols. Malicious data was generated through both network-based (e.g., DoS and TCP port scanning) and host-based attack scenarios. The three datasets are available in CSV format with extracted features.

\begin{table}[H]
\centering
\caption{Dataset samples distribution}
\begin{tabular}{@{}lll@{}}
\toprule
& \textbf{Network} & \textbf{HPC/Kernel} \\ \midrule
\textbf{Benign}         & 2000   & 32303\\
\textbf{DoS}       & 65790   & 65790\\
\textbf{Recon}       & 65790   & 65790\\ \bottomrule
\end{tabular}
\label{tab:distSamp}
\end{table}

%************************************************

\subsection{Experimental Results}

The system was trained and tested in Google Colab using PyTorch for local and federated models. An autoencoder condensed the data into 32 key features via its bottleneck layer. For classification, we used a 1D CNN with two convolutional layers (each followed by max-pooling) and a dense layer. Training employed the Adam optimizer, categorical crossentropy loss, and accuracy metrics. Key CNN hyperparameters are in Table \ref{tab:CNNparam}.

To evaluate detection performance, we measured the false positive rate (FPR) alongside key metrics: accuracy, precision, recall, and F1-score.

\subsubsection{Separate vs. Fused Models}
\begin{figure}[htbp]
    \centering
    \begin{tikzpicture}
        \begin{axis}[
            width=\linewidth, % Make it fit within the column width
            height=6cm, % Adjust height as needed
            ybar=2pt,
            yticklabel={\pgfmathprintnumber{\tick}\%},
            enlarge x limits=0.2,
            ylabel={\%},
            symbolic x coords={Accuracy, Precision, Recall, F1-Score},
            xtick=data,
            legend style={at={(0.5,-0.15)}, anchor=north, legend columns=-1},
            nodes near coords,
            every axis plot/.append style={thick,mark=none},
            x tick label style={rotate=0}, % Rotate x-axis labels for better fit
            ymajorgrids=true
        ]
        \addplot coordinates {(Accuracy,92.91) (Precision,92.94) (Recall,92.91) (F1-Score,92.65)};
        \addplot coordinates {(Accuracy,92.21) (Precision,92.04) (Recall,92.21) (F1-Score,90.15)};
        \addplot coordinates {(Accuracy,90.54) (Precision,90.79) (Recall,90.54) (F1-Score,90.61)};
        
        \legend{Fusionned, Network, Kernel}
        \end{axis}
    \end{tikzpicture}
    \caption{Comparison of performance metrics for Fusionned, Network, and Kernel models.}
    \label{fig:performance_comparison}
\end{figure}
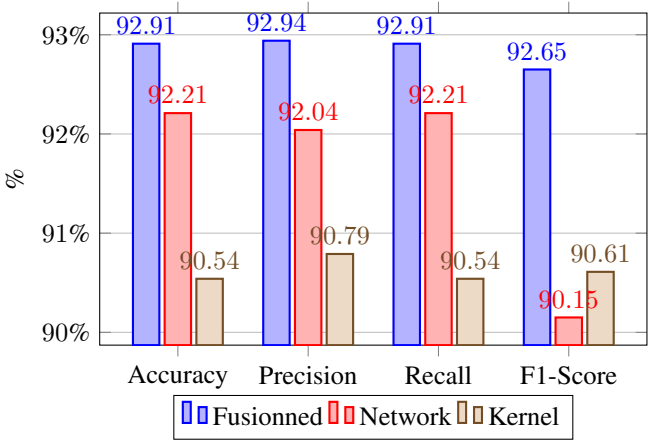

We evaluated detection performance by comparing single-modal models, focused on network traffic and kernel data, with our multimodal approach, which fuses both data sources. A local comparative experiment (excluding Federated Learning) was conducted to assess the impact of data fusion on performance. The results, presented in Figure \ref{fig:performance_comparison}, show that the fusion-based model achieves superior detection performance with an accuracy of 92.91\%, compared to 92.21\% for the network-based model and 90.54\% for the kernel-based model. Additionally, the fusion approach outperforms single-modal models across other metrics, including Precision and F1-Score. The improvement in performance, highlights the effectiveness of using a single, unified model that outperforms models trained on separate data sources.

%************************************************
\begin{table}[H]
\centering
\caption{CNN Model Configuration and Hyperparameters}
\label{tab:CNNparam}
\begin{tabular}{@{}ll@{}}
\toprule
\textbf{Parameter}         & \textbf{Description}                                                                                      \\ \midrule
\textbf{Model Type}        & 1D Convolutional Neural Network (CNN)                                                                        \\
\textbf{Loss Function}     & categorical\_crossentropy                                                                                \\
\textbf{Evaluation Metric} & accuracy                                                                                                  \\
\textbf{Optimizer}         & adam                                                                                                      \\
\textbf{Epochs}            & 10                                                                                                        \\
\textbf{Batch Size}        & 32                                                                                                        \\
\textbf{Aggregation}       & Weighted Average                                                                                          \\ \bottomrule
\end{tabular}
%\end{adjustbox}
\end{table}

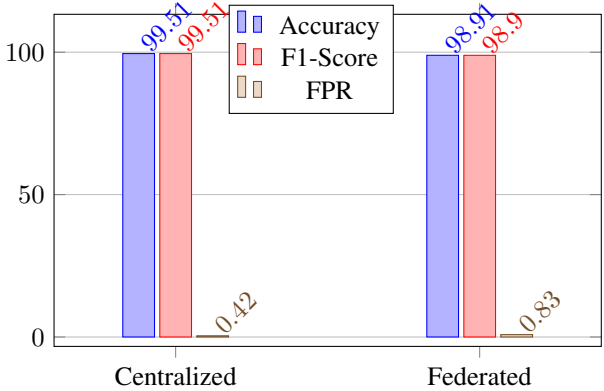
\begin{figure}[htp]
\centering
\pgfplotsset{width=\linewidth,height=6cm,compat=1.16}
\begin{tikzpicture}
\begin{axis}[
      ybar,
    enlarge x limits=1,
    ylabel={},
    symbolic x coords={Centralized, Federated},
    xtick=data,
     enlarge y limits={upper,value=0.2},
    nodes near coords,
    enlarge y limits={rel=0.2,upper},
    ybar,
    legend style = {
     at = {(.47,.7)},
	 anchor=south,
	 column sep = 1ex
	 },
    every node near coord/.append style={rotate=45, anchor=west},
    bar width = 12pt,
    enlargelimits=0.03,
    x tick label style={align=center},
    enlarge x limits=0.4,
    ymax=110,
    bar width=12pt,
        %height=.5\textwidth,
        ymin=0,
        ymajorgrids = true,
    ]  				
\addplot coordinates {(Centralized,99.51) (Federated,98.91)};

\addplot coordinates {(Centralized,99.51) (Federated,98.90)};

\addplot coordinates {(Centralized,0.42) (Federated,0.83)};

\legend{Accuracy,F1-Score,FPR}
\end{axis}
\end{tikzpicture}
\caption{Performances comparison between centralized and federated learning}
\label{bar:comp}
\end{figure}

%*******************************************

\begin{table*}[h!]
\vspace*{0.2in}
\caption{Experimental Results}
\centering
\resizebox{\textwidth}{!}{%
\begin{tabular}{lcccccccccc}
\toprule
& CS 1 & CS 2 & CS 3 & CS 4 & CS 5 & CS 6 & CS 7 & CS 8 & CS 9 & CS 10 \\
\midrule
\textbf{Centralized} & \textbf{Acc: 99.38} & \textbf{Acc: 99.79} & \textbf{Acc: 100.00} & \textbf{Acc: 99.59} & \textbf{Acc: 98.77} & \textbf{Acc: 99.59} & \textbf{Acc: 99.79} & \textbf{Acc: 99.38} & \textbf{Acc: 99.18} & \textbf{Acc: 99.59} \\
& Pre: 99.40 & Pre: 99.80 & Pre: 100.00 & Pre: 99.62 & Pre: 98.76 & Pre: 99.59 & Pre: 99.80 & Pre: 99.39 & Pre: 99.21 & Pre: 99.60 \\
& Rec: 99.38 & Rec: 99.79 & Rec: 100.00 & Rec: 99.59 & Rec: 98.77 & Rec: 99.59 & Rec: 99.79 & Rec: 99.38 & Rec: 99.18 & Rec: 99.59 \\
& F1: 99.39 & F1: 99.80 & F1: 100.00 & F1: 99.60 & F1: 98.75 & F1: 99.59 & F1: 99.80 & F1: 99.38 & F1: 99.18 & F1: 99.59 \\
\midrule
\textbf{Federated} & \textbf{Acc: 98.97} & \textbf{Acc: 98.77} & \textbf{Acc: 98.97} & \textbf{Acc: 98.36} & \textbf{Acc: 98.97} & \textbf{Acc: 98.97} & \textbf{Acc: 98.77} & \textbf{Acc: 98.97} & \textbf{Acc: 98.77} & \textbf{Acc: 98.97} \\
& Pre: 98.98 & Pre: 98.80 & Pre: 99.00 & Pre: 98.38 & Pre: 98.97 & Pre: 98.98 & Pre: 98.87 & Pre: 98.98 & Pre: 98.88 & Pre: 98.98 \\
& Rec: 98.97 & Rec: 98.97 & Rec: 98.98 & Rec: 98.36 & Rec: 98.97 & Rec: 98.97 & Rec: 98.87 & Rec: 98.97 & Rec: 98.77 & Rec: 98.97 \\
& F1: 98.97 & F1: 98.97 & F1: 99.00 & F1: 98.37 & F1: 98.96 & F1: 98.97 & F1: 98.76 & F1: 98.97 & F1: 98.87 & F1: 98.97 \\
\bottomrule
\end{tabular}%
}
\label{table:results}
\end{table*}

\begin{table}[h!]
    \caption{Performance Metrics by Number of Clients}
    \centering
    \begin{tabular}{@{}lcccc@{}}
        \toprule
        \textbf{Nb. CS} & \textbf{Accuracy (\%)} & \textbf{F1-Score (\%)} & \textbf{False Positive Rate ((\%)} \\ 
        \midrule
        \textbf{10}          & 98.91            & 98.90             & 0.83         \\ 
        \textbf{8}           & 98.64            & 98.63             & 1.52         \\ 
        \textbf{6}           & 98.89            & 98.88             & 0.67         \\ 
        \textbf{3}           & 99.61            & 99.61             & 0.40         \\ 
        \bottomrule
    \end{tabular}
    \label{resultsByClients}
\end{table}
%****************************************

\subsubsection{Centralized vs. federated}
We compare the results for both the centralized and federated models across 10 Charging Station (CS). The results are shown in Table \ref{table:results} and Fig. \ref{bar:comp}.
%and Fig. \ref{fig:comparison_results}.
%******************************

%*******************************

%\subsubsection{Centralized Results}
The centralized model achieves near-perfect performance across all clients, with metrics often exceeding 99\%. For instance, Client 3 recorded 100\% in accuracy, precision, recall, and F1 scores, highlighting its ability to generalize effectively by leveraging a comprehensive, centralized dataset. The federated model shows a slight performance drop, with accuracy ranging from 98\% to 99\%, such as 98.36\% for Client 9. Despite this, it maintains strong detection rates while preserving data privacy, striking a practical balance between accuracy and decentralization.

\subsection{Federated Performance Across Client Numbers}
The results in Table \ref{resultsByClients} illustrate that the federated learning model maintains robust performance even as the number of clients increases. With 10 clients, the model achieves an impressive accuracy of 98.91\% and an F1-score of 98.90\%, with a low false positive rate (FPR) of 0.83\%. This consistency across metrics, regardless of client count, underscores the model’s capacity to handle distributed data effectively without compromising accuracy. While the FPR sees a slight uptick with more clients, the model continues to deliver reliable results, affirming its scalability and adaptability across diverse data sources while retaining high accuracy and minimal false positives.

\subsubsection{Discussion}
The results show that, although the centralized model slightly outperforms the federated approach for some clients, the federated framework still achieves performance levels close to those of the centralized model. The small performance differences can be attributed to local variations in data distribution, which tend to be more pronounced in federated settings. Nonetheless, the federated model proves to be an effective and practical solution, striking a balance between privacy and detection accuracy across a distributed system. These findings highlight the potential of federated learning to maintain high accuracy while ensuring that individual EVSEs retain control over their own data.

\section{Conclusion and Future Work}
We propose a federated multimodal IDS for Electric Vehicle Supply Equipment (EVSE). Our findings highlight the effectiveness of combining data fusion with federated learning, enhancing intrusion detection across EVSE networks. By integrating multiple data sources, our framework enables comprehensive threat detection with high accuracy. The federated approach keeps data decentralized, ensuring privacy without compromising performance. This combination provides a scalable, secure solution for EVSE infrastructure in the smart grid ecosystem. Future work will optimize model efficiency, integrate additional data modalities, and explore real-time deployment in more EVSE scenarios. We will also address FL robustness against poisoning attacks through secure aggregation techniques.

\section*{Acknowledgment}
This work is supported by the OPEVA project that has received funding within the Chips Joint Undertaking
(Chips JU) from the European Union’s Horizon Europe Programme and the National Authorities (France, Czechia, Italy, Portugal, Turkey, Switzerland), under grant agreement 101097267. In France, the project is funded by BPI France under the France 2030 program on "Embedded AI". Views and opinions expressed are however those of the author(s) only and do not
necessarily reflect those of the European Union or Chips JU. Neither the European Union nor the granting authority can be held responsible for them.

\bibliography{ref}

\end{document}